# Cyber Security and Applications
## AI-Driven Security in Cloud Computing: Enhancing Threat Detection, Automated Response, and Cyber Resilience
--Manuscript Draft--

| | |
|---|---|
| relevant data files) as a single zip file, and select "Data in Brief" as File Type. | |
| **Free Preprint Service**<br><br>Do you want to share your research early as a preprint? Preprints allow for open access to and citations of your research prior to publication.<br><br>Cyber Security and Applications offers a free service to post your paper in a journal-branded First Look space on SSRN, an open access research repository, when your paper enters peer review. Once on SSRN, your paper will benefit from early registration with a DOI and early dissemination that facilitates collaboration and early citations. It will be available free to read regardless of the publication decision made by the journal. This will have no effect on the editorial process or outcome with the journal. Please consult the SSRN Terms of Use and FAQs. | YES, I want to share my research early and openly as a preprint. |

# AI-Driven Security in Cloud Computing: Enhancing Threat Detection, Automated Response, and Cyber Resilience

Shamnad Mohamed Shaffi
Data Architect
Amazon Web Services
Seattle, WA, US
shamnadshaffi@ieee.org

Sunish Vengathattil
Sr. Director, Software Engineering
Clarivate Analytics
Philadelphia, PA, USA
sunish_v_nair@ieee.org

Jezeena Nikarthil Sidhick
Senior Data Engineer
American Express Global Business Travel
Bellevue, WA, USA
jezeena@gmail.com

Resmi Vijayan
Software Engineer
Comcast
Philadelphia, PA, USA
resmi_vijayan@comcast.com

*Abstract:* Cloud security concerns have been greatly realized in recent years due to the increase of complicated threats in the computing world. Many traditional solutions do not work well in real-time to detect or prevent more complex threats. Artificial intelligence is today regarded as a revolution in determining a protection plan for cloud data architecture through machine learning, statistical visualization of computing infrastructure, and detection of security breaches followed by counteraction. These AI-enabled systems make work easier as more network activities are scrutinized, and any anomalous behavior that might be a precursor to a more serious breach is prevented. This paper examines ways AI can enhance cloud security by applying predictive analytics, behavior-based security threat detection, and AI-stirring encryption. It also outlines the problems of the previous security models and how AI overcomes them. For a similar reason, issues like data privacy, biases in the AI model, and regulatory compliance are also covered. So, AI improves the protection of cloud computing contexts; however, more efforts are needed in the subsequent phases to extend the technology's reliability, modularity, and ethical aspects. This means that AI can be blended with other new computing technologies, including blockchain, to improve security frameworks further. The paper discusses the current trends in securing cloud data architecture using AI and presents further research and application directions.

## I. INTRODUCTION

Cloud computing has altered how enterprises store and process their data, making it possible to expand the solutions while simultaneously containing high overheads. Since transcended beyond implementation in academic and research establishments, it has surmounted significant breakthroughs in data sharing, productivity, and use [1]. Nevertheless, growing reliance on cloud technology has led to new threats, such as hackers, thefts, and internal threats, which are threats to organizations [3]. This means that traditional security firewalls and encryption approaches are not enough to counter present and future threats [2].

AI is the new trend in cloud security since it is a powerful tool for detecting threats, monitoring for anomalies, and responding automatically. Security solutions now use machine learning and predictive analytics to prevent a threat that is likely to be damaging from practically occurring [3]. As AI processes big data from the cloud and eventually learns it, it improves the security position, shortens reaction times, and decreases human mistakes, thus becoming a crucial aspect of modern-day protective measures [6].

This paper seeks to discuss the use of AI-supported solutions in the context of cloud data architecture and the opportunities, challenges, and prospects of this approach. It reviews prior research studying AI in cybersecurity, explores issues like data privacy and regulatory concerns, and presents an outlook on the potential means of improving cloud security AI. The study expects to establish findings that will enlighten how adopting AI changes the cloud security environment and prepares the environments for better security.

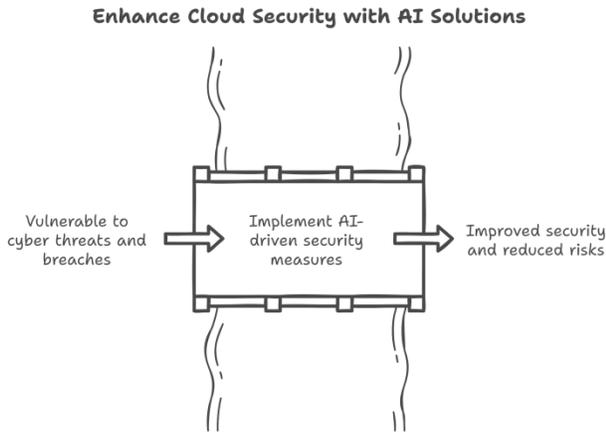

Fig 1: Enhance Cloud Security with AI Solutions

## II. LITERATURE REVIEW

Cloud computing has become a burgeoning technology because of the rapid increase in the storage and use of data by individuals and companies. Attackers keep launching new attacks into cloud environments to target clients, gaining unauthorized access to their data and leading to data leakages and compromised cloud services.

New threats have occurred, highlighting the problems of cloud security, which require further strengthening of security measures. The increased utilization of cloud platforms weakens organizations' assets' links to cyber threats, as current threats reveal whether current security is enough.

The most recent was in July 2024, when a Crowdstrike security breach affected several organizations worldwide. The matter involved CrowdStrike's failure to secure its endpoint detection and response (EDR) system, whereby the security measures were compromised, and the adversaries got access to the information they were not supposed to access. These problems axed all aspects of governance and financial and healthcare facilities. The source of this attack was realized to have originated from a vulnerability that was not patched to allow the attackers to penetrate the systems. This breach prompted the discussion about the third-party cybersecurity vendors' dependency and proactivity of security patching [1].

Another one was a cogent data leakage vulnerability in Microsoft Azure. This led to cloud misconfiguration, poor identity and access management policies, and unauthorized access to several enterprise data information. The attacker used misconfigured storage accounts and unsafely protected APIs to steal sensitive data. This incident supports the accumulation of threats related to misconfigured cloud services, including cloud leaders. The incident impacted businesses that depend on Azure services and artificially forced customers to rethink the Azure shared responsibility model [2].

Another recent and well-known case of corporate hacking happened to AT&T; the company lost its consumer data. The attackers demanded that they be paid in Bitcoin to remove the stolen data; the experience unveiled a danger of ransom based on threats. Millions of AT&T clients' data were accessed, including personal identification details, account credentials, and billing information. The core of the problem was found to lie in AT&T's data management and security, which allowed the customer data to be extracted easily. As for significant points of the described incident, it pointed to the possibility of suffering considerable monetary and image losses due to computer break-ins and increased activity of cybercriminals who tend to turn to ransom demands as an effective way to leverage data stolen [3].

These are some of the reasons why higher levels of security are required for those offered by encryption and firewalls. They stress applying AI solutions to protect an organization's assets from threats, secure against an improper configuration, or even enhance end-point protection. Focusing on the future, threat identification, automation of security, and monitoring procedures represent the crucial components of protecting cloud environments.

Various conventional protection tools exist, like firewalls, encryption, and antivirus programs, but they do not suffice enough to protect from the new, improved types of cyber threats. Thus, AI has become one of the most effective concepts in addressing cloud security issues and providing innovative, advanced procedures for protection. This conceptual background section offers a synthesis of cloud security issues starting from current third-party literature regarding familiarization with security constraints, traditional security models, and innovative AI-based cybersecurity risk management solutions generated and supported through realistic case studies of AI mitigation of threats.

*A. Preliminary Survey Of Threats On Cloud Environment: Key Threats And Their Impact*

The problems of cloud security are most evident when threats of the threat actors are exploited to attack the cloud infrastructure of businesses. Risks include data loss, DoS attacks, data leak insider attacks, and ransomware attacks. Research has illustrated that up to 80% of cloud breaches are caused by mistakes and poor access control [4]. Other forms of social engineering, such as phishing, also play a role in credential theft, hence unauthorized access to information. Advanced persistent threats (APTs) operate in cloud environments where the attackers' presence goes unnoticed for a long time, and the long-term goal is acquiring essential data [5]. One of the burning issues from the further enhancement of the next generation of computing environment, namely, multi-cloud and hybrid cloud, is security issues and threats or the need to transition from post-factum protection to the concept of prevention.

*B. Limitations of Traditional Cloud Security Measures: Why Conventional Approaches Are No Longer Sufficient*

The foundation of cloud protection includes conventional cloud security tools, such as firewalls, encryption mechanisms, IDSs, and antivirus software, which have significant limitations. However, traditional security systems are between real-time adaptable and thus not a suitable technique for dealing with portable and developing cyber threats in today's computing environment [6]. Furthermore, these tools are based on previously identified attack patterns; hence, they cannot detect and neutralize zero-day threats [7]. Furthermore, many conventional security solutions rely on manual intervention, which slows down threat containment and increases the probability of success of cyberattacks [8]. Moreover, as cloud adoption continues to increase, these old-school security ways become more challenging to scale as they struggle to handle vast amounts of traffic and distributed data, and these organizations are very vulnerable to security attacks.

*1) Incident Response (IR) Challenges and AI/ML Solutions*

Cloud security with incident response (IR) has many challenges in dealing with the complex nature of cyber threats and distributed cloud environments. The main problem comes from being able to detect threats in real-time in multiple cloud infrastructures. These traditional security tools find it challenging to correlate security events on the different platforms, resulting in delayed fusion of security events for threat detection and response. In addition, cloud environments show high sophistication and evolution in attack patterns that do not allow rule-based security systems to change in practical ways entirely. Unsurprisingly, security teams also see an avalanche of alerts daily, most of which are false positives, resulting in alert fatigue and delaying an appropriate response to real threats. On the other hand, cloud assets are not easily visible to organizations, which means it is difficult for them to discover unauthorized access or malicious activities and fail to discover them before they escalate into cyber break-ins [13].

In order to tackle these challenges, AI and ML-driven solutions have transformed incident response by providing auto security operations and predictive threat mitigation. The security teams can filter out the false positives, and machine-learned prioritization of alerts will allow the security teams to focus on high-risk threats. Security Orchestration, Automation, and Response (SOAR) platforms based on AI are used to integrate various security tools in one platform by which automated incident containment and response execution are enabled. As described above, deep learning models can enhance threat detection by analyzing massive log datasets and identifying suspicious patterns and anomalies indicative of potential attacks by detecting them.

*2) Shared Responsibility Model Challenges and AI/ML Solutions*

Nevertheless, cloud security needs to be divided, split, and shared between the CSPs and the customers, or instead, the responsibilities must be split in some way for Cloud Security to be executable. However, the flexible division of roles brings some confusion and the possibility of numerous security risks. Governance issues pose a problem mainly due to a lack of definition and respective identification of boundaries of organizations' liabilities, which in turn makes misconfigurations and exposures widespread. In this context, because workloads constantly move between public, private, and hybrid clouds, security policies have always been challenging to standardize. Moreover, it is crucial to have the support of information technology personnel, security personnel, compliance officers, and executives to enforce security. Other factors that make cloud security management more challenging include regulatory compliance, where policies like the GDPR, HIPAA, and ISO 27001 need constant monitoring and change to prevent costs of compliance and legal suits [9].

These challenges can be addressed through AI and ML solutions designed to incorporate intelligence in security monitoring to map out security responsibilities. Machine learning algorithms review cloud configurations to determine potentially dangerous structural vulnerabilities and certify that the organization complies with its regulatory standards. Businesses cut down the amount of time spent and intervention by automated means of checking compliance by comparing the cloud infrastructure against specific regulations. The risk assessment tools allow AI to expose potential threats to an organization before they are exploited, thus improving their prevention. Furthermore, AI offers security posture management that allows cloud settings to stay valid according to the current best practices and risk modeling that forecasts risk based on its data regarding past attacks. Using AI, the business can manage and facilitate defined communicational interactions between the involved stakeholders and specific task assignments and have an innovative approach to workflow, which may enhance security significant response speed [10].

*3) Data Protection Challenges and AI/ML Solutions*

Businesses have a serious problem with protecting sensitive data in the cloud because of the complex encryption requirements, the ever-evolving data privacy regulations, and the decentralization of cloud storage. To protect data, organizations must implement robust encryption mechanisms such that unauthorized users can access only the data they can. While managing encryption keys manually comes with vulnerabilities, it is still possible to do so. Since securing data privacy and enforcing strict security policies on a company must comply with data protection laws such as GDPR, CCPA, and PCI DSS, it is important to keep an eye on your system and configure and implement regular audits. What makes it worse is that multiple data storage locations across various cloud environments make it challenging to have a central eye on security. Unauthorized access, insider threats, and misconfigured access controls increase the likelihood of a data breach, which, among other potential losses, results in a loss of money and damage to reputation [11].

The challenge of data protection is solved by using AI and ML-driven solutions to automate encryption management and enforce compliance policy. The machine learning algorithm provides a way of automatically allocating the encryption keys based on data sensitivity and reduces the manual intervention and the security gaps. This means that AI-powered access control systems monitor each user's behavior. They will likely engage in suspicious activity, such as unauthorized access attempts, whenever they detect strange behavior. By intelligent data classification, AI systems can classify data on the risk level and assign a different level of protection for highly sensitive information. Moreover, AI-driven solutions for privacy protection, including AI-driven PII detection and AI-powered data masking, ensure compliance and maintain the customer's information. Data loss prevention (DLP) solutions based on AI analyze network traffic and prevent possible data leaks by detecting unauthorized transmission. This allows cloud-based data to be secure while promoting changes to regulatory requirements [12].

Cybersecurity has become a critical issue in organizations because malicious attacks have advanced more than ever, requiring organizations to develop intelligent, automatic, and adaptive security capable of detecting and responding to such threats in real time.

*C. AI-Based Approaches to Cybersecurity*

*1) Machine Learning for Anomaly Detection in Cloud Security:* Applying ML models to examine large volumes of data and determine if a potential cyber-attack exists is possible. However, ML algorithms learn and evolve with time compared to conventional rule-based security systems. Supervised learning models detect known threats, such as decision trees and support vector machines. In contrast, unsupervised learning models, such as clustering algorithms, detect unknown threats based on deviation from normal network behavior [9].

*2) Deep Learning for Identifying Hidden Threats in Cloud Networks:* A deep learning approach uses artificial neural networks to identify mixed patterns in a data set. It is also very good at identifying zero-day threats and other specific types of malwares. Besides, it may detect hazardous viruses that can avoid detection by standard anti-virus software. Network traffic logs, alerts, and endpoint behavior used with deep learning greatly help reduce false positives in threat detection [10].

*3) Behavioral Analytics for Proactive Intrusion Detection and Risk Assessment:* To find these patterns, supervised behavior analysis tracks the user's activities, device interactions, and network traffic. With the set normal behavior, the AI systems can capture eventualities that depict the account's anomalous status, unauthorized access, or even infection by malware [11]. This can be useful for various businesses since this action is preventive and ensures that security threats that may threaten the organization's data do not occur.

*D. Case Studies of AI-Driven Security Implementations: Real-World Applications and Success Stories*

Integrating AI-driven security models in cloud systems has made a massive shift in data safety, threat identification, and response time. More than one organization, irrespective of the type of business, has implemented AI-based security frameworks to prevent cyber-attacks and protect data. The following case studies reflect real-life cases where the use of AI-based security systems has helped boost the security system.

*1) AI-Driven Zero-Trust Architecture (ZTA) in Large Enterprises*

ZTMAc is a cybersecurity architectural model that does not inherently trust any subject, internal or external to the network. Instead, it implements strong identity checks and constant authentication features. An attempted case was developed by a large financial institution where using an ML model, they built a Zero-Trust security structure that would monitor user behavior with high frequency, analyze access requests, and alert and prevent any anomalous behavior in real time. The results were remarkable:

- Alone, continuous authentication and real-time anomaly detection made with AI can reduce unauthorized attempts by 85%.

- There was a much higher decline in insider threats since the AI agent was able to monitor internal operations and identify suspicious access patterns that may pose threats to insiders.

- Reducing the likelihood of human error, enforcing least privilege access, and making access decisions automatically.

The following makes it evident that the use of AI in ZTA improves security because it disapproves of any implicit trust while amplifying the security measures according to threat intelligence collected in real-time [12].

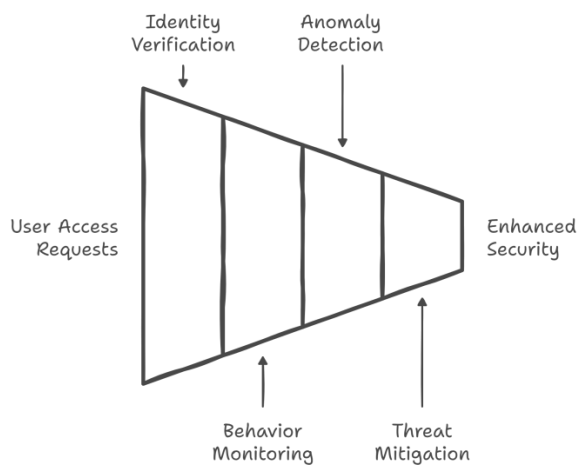

Fig. 2: AI-Enhanced Security in Zero-Trust Architecture

*2) Artificial Intelligence in SIEM: The Ways to Strengthen Real-Time Analyzing and Threat Information*

SIEM systems are imperative solutions for assembling, evaluating, and correlating information about security occurrences from multiple sources to identify cyber threats in cloud networks. However, the traditional approaches implemented in SIEM are ineffective because of challenges. This is because traditional SIEM solutions have false positives, slow response time, and are incapable of coping with massive data logs.

The following benefits have, therefore, been realized by companies that have implemented AI-driven SIEM solutions:

- Event correlation and threat detection of up to 60% could be achieved without spending considerable time on this process and interacting with cyber threats.

- Improved efficiency in threat prioritization, as AI models prioritize security alerts in an optimal order to evade overwhelming security staff with too many alerts.
- Changes detection involves Feed Forward, where higher forms of threat intelligence learn from security threats to refine future risk identification.

Now, with the integration of the advanced components of Artificial Intelligence, SIEM has become more proactive and has made it easier for the security teams to work more effectively and respond to threats and incidents with high levels of precision [13].

*3) Deep Learning for Fraud Detection in E-commerce Platforms*

Because the internet is available and accessible to customers, credit card fraud via cyber scams, threatened chargebacks, stolen credit cards, and the creation of fake accounts is relatively common in online retailing companies. Analytical fraud detection methods tend to have updated problems with new fraud behaviors or produce too many false alarms, disturbing legitimate customers' transactions. One of the largest online selling stores decided to establish a deep learning-based fraud detection system, which had the following benefits:

- By achieving an accuracy rate of 97% when determining fraudulent transactions, chargebacks and financial losses are minimized [14].
- Real-time risk mitigation, where the AI in an environment constantly assesses the purchase behaviors, the fingerprints of the device being used, and past transactions to identify fraudulent activity on the fly.
- Reducing and avoiding itself after allowing decision-making while ensuring that only fraudulent transactions are prevented.

The application of deep learning here can be described as the presence of a learning algorithm that captures new forms of fraud detection, recognizes complex schemes, and improves the overall security of the transactions [14].

TABLE 1: DEEP LEARNING FRAUD DETECTION

| Deep Learning for Fraud Detection in E-commerce Platforms | Details |
|---|---|
| Problem | E-commerce forms include credit card fraud and chargebacks, stolen credit cards, fake accounts, and cyber fraud. Traditional fraud detection techniques are inadequate to deal with new fraud scenarios and generate large numbers of false positives. |
| Solution | Deciding on improving the fraud detection system using a deep learning method to increase the fraud detector's effectiveness. |
| Key Benefits | - 97% efficiency in the cases of fraudulent transactions, chargebacks, and, consequently, financial losses. |
| | - Risk administration in real-time by analyzing user behavior, identifying a device, and tracking transaction history. |
| | - It also lowers false positives, eliminating cases where valid transactions or operations are denied. |
| Impact | It learns new fraud patterns, can identify other types of fraudulent scenarios, and increases transaction protection. |

*4) AI-Enhanced Threat Hunting and Response in Cloud Data Centers*

Business-critical data stored in cloud data centers is promising for APTs, insider attacks, and ransomware threats. Most cloud service providers have adopted AI algorithms in their security models to increase the chances and mechanisms of active threat searching and prevention. Some of the notable advantages of integration of artificial intelligence in cloud security include the following:

- Ongoing monitoring and identification of behaviors indicative of APTs are done by utilizing an automated detection system on the network traffic.
- Real-time prevention of ransomware attacks where files infected by viruses are shut down, unauthorized access is prevented, and encryption of other files is also halted.
- Intelligent security features allow security teams to learn about some kinds of threats before they become imminent threats.

Thus, by using AI for constant threat detection, prediction, and response, CS teams can minimize the time needed to address incidents and improve the overall security of cloud ecosystems [15].

*5) Comparative Analysis of AI Models in Cloud Security: Strengths and Weaknesses of Different AI Techniques*

There are various aspects concerning the AI models applied to cloud security and the strengths and disadvantages of specific models. The following table gives a cross comparison:

TABLE 2: STRENGTH AND WEAKNESS OF DIFFERENT AI MODELS

| AI Models | Advantages | Limitations |
|---|---|---|
| Supervised Learning | The ability to detect known threats is highly accurate | Requires labeled datasets for training |
| Unsupervised Learning | Identifies unknown threats and anomalies | May generate false positives |
| Reinforcement Learning | Continuously improves defense mechanisms | Requires extensive computing resources |
| Deep Learning | Excels at recognizing zero-day attacks | High computational cost |
| Behavioral Analytics | Several of its capabilities include real-time | Requires further enhancement to reduce the number of alarms |

| | identification of inside threats | |

These methods bring the best and most progressive cloud security options parallel to the next level of AI models.

The increasing sophistication of cyber threats necessitates a paradigm shift in cloud security strategies. Security solutions incorporating Artificial Intelligence are proficient in predetermination, automation, and flexibility of Security Algorithms as opposed to conventional security techniques. Nevertheless, relative risks, including adversarial AI attacks, data privacy, and computational costs, must be overcome to leverage AI to advance cloud security fully.

Research should be conducted in the future to further improve AI's interpretability, develop integration with blockchain, advance threat intelligence, and develop next-generation cloud security using AI.

## III. FUTURE WORK

With cyber threats growing increasingly complex, it is not surprising that AI-based security solutions must adapt to start fighting these very complex attacks. The future of cloud security is emerging AI technologies, improved threat intelligence systems, ethical considerations in the case of AI-powered security, and integration with other innovative technologies such as blockchain. Due to the increasing use of AI to standardize and increase the effectiveness of cybercriminal attacks, cybersecurity professionals must adopt AI-based solutions to detect, predict, and avert threats proactively. Within this section, details are given on where future research and development will take the next generation of cloud security.

*A. Emerging AI Technologies for Enhanced Cloud Security*

Like any new technology, AI is rapidly advancing to new frontiers in cloud security, particularly more innovative, adaptive, and autonomous defense mechanisms. The quantum integration of quantum machine learning (QML) can change cloud security by adding quantum computing's power to analyze big data in real time, revolutionizing the cloud security market. This will consequently significantly improve the detection of the zero days that typically evade conventional security measures. This is another promising development used in security, specifically threat simulation and using GANs to train AI security models against ever-changing cyber threats. The utility of GANs is to generate sophisticated attack scenarios to improve security systems' resistance against adversarial attacks.

Another is that autonomous AI security agents are created to act against threats without human intervention, and cybersecurity becomes more proactive than reactive. These agents will use reinforcement learning techniques to adapt to new cyber threats on the fly so that response times are reduced, and the impact of a breach is minimized. Also, in the future, AI will come up with self-healing AI systems capable of automatically detecting and patching vulnerabilities and thus preventing a security incident from escalating. Another steeping field where AI in automating compliance management has grown rapidly is through intelligent systems monitoring regulatory requirements and ensuring that the cloud security measures align with evolving global cybersecurity standards.

*B. Advancements in AI-Driven Threat Intelligence*

New technologies in cloud security will emerge in successive variations of intelligent threat intelligence where security systems are aligned more to anticipation and prevention than reaction. The evaluation of the threat source is expected to convert to more innovative algorithms, where cyberattacks can be tracked to their sources. These models will require deep learning to analyze data from several sources, recognize the profile of attacks, and estimate future attacks.

Besides the feature of attribution of an event, context-aware anomaly detection is the next big step toward AI-driven threat intelligence, which adds behavioral analytics, device fingerprints, and network activity to identify anomalous behaviors of a person in real time. In contrast to other anomaly detection systems based on the rules of logical decision-making, AI models will improve with each new data and work accurately to detect an insider threat or a complicated cyberattack. Another relatively explored area is federated learning, which is training a learning model in multiple organizations while sharing the security data analyses without exchanging actual data. It is a decentralized model that will help to enhance global protection against threats and make organizations more effective in analyzing threats.

Future developments will also establish a link between AI and cyber threat intelligence feeds that collate information from the government, cyber security companies, and organizations. These platforms will assimilate AI to offer threat intelligence to organizations and help organizations prevent such threats by taking measures even before they are hatched. At the same time, advanced technologies such as honeypots and traps will be employed in security systems to deceive attackers and gain data on their TTPs.

*C. Ethical and Privacy Considerations in AI-Powered Security*

This is because the companies' use of AI for cloud security poses several ethical and privacy issues. It is also possible to prioritize security threats based on prejudices we have of potential attackers and thus create a flawed security model. This is because, with trained AI models that use existing data, there might be more focus on specific types of user behavior while some elaborate cyber-attacks are left unnoticed. This can be done with the help of collecting data that is diverse and inclusive about gender, race, etc., and using specific algorithms that were developed with fairness in mind. The next concern is the problem of surveillance that is put in place by AI monitoring solutions since they constantly track various user activities. Open AI governance policies and the use of applications, including differential privacy and homomorphic encryption, should regulate artificial intelligence in organizations.

Besides, artificial intelligence is a sensitive element in cyber defense; therefore, accountability and explainability of its decisions are crucial. Thus, these systems should be explainable and accountable when artificial intelligence is given specific security responsibilities. This will bring about explainable AI (XAI), which will aid the cybersecurity teams in comprehending how the models identify threats and their choices to minimize numerous ignorance issues and boost the trust and utilization of AI security systems. They will have to define regulations regarding the proper use of AI in cloud security, and governments and industries will do what is needed. Accountability in artificial intelligence and adherence to the laws governing data privacy are other factors that will explain the general acceptance of AI in security systems.

*D. AI and Blockchain Integration for Next-Generation Cloud Security*

In order to enhance cloud security, integrating AI with blockchain offers an innovative way to integrate AI's predictive capability with blockchain's decentralized and tamper-proof architecture. The creation of blockchain identity management with assistance from AI can enhance authentication by authenticating user identities while ensuring that the information is immutable and tamper-proof. This integration can significantly cut identity fraud and unauthorized access to cloud resources. Further, the AI-driven security models can use consensus mechanisms in blockchain to validate the threat intelligence in a distributed network with enhanced accuracy and reliability of cyber threat detection.

Blockchain can also help improve the transparency and accountability of AI-driven security measures. It will benefit industries where data protection is crucial and necessary, such as healthcare and finance. Also, with the help of AI-powered smart contracts, AI can regulate and automate security policies, incident response, and compliance enforcement, making cloud security proactive and efficient. With the evolution of AI and blockchain technologies, the use case of securing cloud infrastructures with emerging cooperation between AI and blockchain will further enhance threat mitigation and trust-building capabilities in cloud-based services.

That is why cloud security in the future will depend on the developments of Artificial intelligence and its interconnectivity with other novelties. Given the constantly increasing cyber threats, IT security solutions that utilize artificial intelligence should be predictive, adaptive, and independent. Advanced quantum machine learning and self-autonomous currents will improve threat detection, especially threat sophistication. Developments in AI in threat intelligence will enable an organization to detect threats and respond to them promptly to prevent them from becoming out of control. Ethical and privacy concerns must surface and be managed to advance the proper use of AI. At the same time, the marriage between AI and blockchain will enable new forms of security layers that are almost immutable to hacker attacks.

## IV. DISCUSSION

Enhancing cloud security with AI has provided better real-time threat detection, early identification of threats, risk prevention strategies, and risk management. Artificial intelligence security technologies apply machine learning algorithms, deep learning algorithms, and behavioral analysis to prevent cyber threats from happening. However, some issues related to such solutions include computational complexity, ethical issues, and regulation issues. Also, AI is implemented differently depending on the cloud environment; the outcomes of the cloud depend on its infrastructures, governance policies, and the type of data being processed in the cloud environment. This section discusses these aspects in further detail in terms of the efficiency, constraints, and conformity of the regulation of AI-powered cloud security.

*A. Effectiveness of AI in Real-Time Threat Detection and Prevention*

AI has brought significant changes regarding the ability to monitor events as they occur. It uses big data to detect corruption and then respond without a human being intervening. Deep learning-based IDS has been used to identify new levels of cyber-attack with adequate security, thus mitigating the risk of data breaches in cloud computing [3]. AI-based threat intelligence also improves the speed of actions since programs are created to fight malware and ransomware, and IT insiders will act without delay [7]. With the help of AI in predictive analytics, an organization can quickly identify areas of weakness and fix them before they are exploited by attackers [12]. In addition, It is noteworthy that Cloud SIEM solutions with AI help to monitor constantly and make cloud platforms stronger against new threats [14].

*B. Limitations and Challenges in Implementing AI-Driven Security Solutions*

Despite its advancements, AI-based cloud security faces several limitations. One of them is that deep learning models require significantly high computational power. Therefore, deploying such models is expensive and computationally demanding [5]. Also, AI programming functions based on data feeding; hence, if the data fed is flawed, the system will be equally flawed, and the results from such a broken algorithm can compromise security measures [9]. Another problem with AI applications is that they are adversarial systems since hackers will also employ AI to hack into the systems by exploiting their programming [11]. There are also various drawbacks regarding the use of automation, such as the assumption by the organization to trust the automated systems, hence making little or no effort to oversee critical security decisions made by the computers as they might misinterpret the alerts stated by the durable AI systems [16].

*C. The Role of Regulatory Compliance and Governance in AI-Based Cloud Security*

AI-operated cloud security must respect regulatory and compliance standards to offer legal data handling. Some examples of mandatory regulation for industries, including the financial and healthcare sectors, are GDPR, HIPAA, and FedRAMP [18]. These laws require implementing measures in security control and data protection and explaining the functions of artificial intelligence [20]. Nevertheless, they also pointed out one of the major problems that organizations experience: rapid changes in the threats within the cybersecurity environment, which, in turn, may result in violations of regulatory requirements when introducing fresh, innovative technologies based on AI [22]. Also, regulatory bodies are trying to adopt AI governance principles that would help manage bias, accountability, and transparency in cybersecurity and enhance the proper use of AI [24].

*D. Comparison of AI Security Frameworks in Different Cloud Environments (Public, Private, Hybrid)*

The security aspects of AI depend on the type of cloud environment adopted in an organization, with different models comprising public, private, and hybrid, which have distinct features in terms of security. AWS, Google Cloud, and Microsoft Azure provide AI, which is integrated into their platforms, and provide solutions that are scalable and economically common but are partially secure due to having shared infrastructure and multi-tenancy [6]. Private cloud services are more customizable and can offer a dedicated data security framework to enterprises dealing with large volumes of confidential information; simultaneously, they are costly as they demand initial investment in AI security solutions [13]. It is important to note that hybrid cloud systems combine both and allow organizations to integrate AI security solutions within diverse environments while considering security and performance or compliance requirements [17]. That said, challenges emerge when integrating security solutions with AI across the hybrid securing condition, meaning these systems must be connected and updated in real-time [19].

In conclusion, there are outstanding issues with computational costs, adversarial threats to AI, and issues of governance surrounding AI-based security solutions despite the solutions helping detect and respond to threats and compliance in the cloud architecture. The future will, therefore, depend on integration, increased regulation, and partnerships between researchers, policymakers, and cloud companies to devise a better way forward.

## V. CONCLUSION

Artificial intelligence has emerged as an important aspect of enhancing security in cloud data structures, given the growing incidence of cyber risk in modern society. AI improves the flow of real-time threat detection, implements prompt incident response, and enhances the security of cloud services, surpassing regular practices at their core through machine learning, deep learning, and behavioral analytics. These points also illustrate how AI decreases the risk factors of cyber threats by allowing the system to forecast weaknesses, minimize human mistakes, and adjust for new patterns. However, issues such as computational requirements, adversarial artificial intelligence, and legal compliance must be solved systematically.

In the future, better development of security models with artificial intelligence, the combination with blockchain technology, and enhancing cloud governance will take cloud security to another level. As AI enhances its use in organizations, it is necessary to balance development and development to foster more secure cloud solutions in delivering organizational services while considering the integrity of services and legal frameworks.

## REFERENCES

[1] Abouelyazid, M., & Xiang, C. (2019). Architectures for AI Integration in Next-Generation Cloud Infrastructure, Development, Security, and Management. International Journal of Information and Cybersecurity, 3(1), 1-19.

[2] Akram, E., & Basit, F. (2023). AI-Powered Information Security: Innovations in Cyber Defense for Cloud and Network Infrastructure.

[3] Ayyadapu, A. K. R. (2023). Enhancing Cloud Security with AI-Driven Big Data Analytics. International Neurourology Journal, 27(4), 1591-1597.

[4] Bolanle, O., & Bamigboye, K. (2019). AI-Powered Cloud Security: Leveraging Advanced Threat Detection for Maximum Protection. International Journal of Trend in Scientific Research and Development, 3(2), 1407-1412.

[5] Dash, B. (2024). Zero-Trust Architecture (ZTA): Designing an AI-Powered Cloud Security Framework for LLMs' Black Box Problems. Available at SSRN 4726625.

[6] Gopireddy, R. R. (2021). AI-Powered Security in Cloud Environments: Enhancing Data Protection and Threat Detection. International Journal of Science and Research (IJSR), 10(11).

[7] HaddadPajouh, H., Khayami, R., Dehghantanha, A., Choo, K. K. R., & Parizi, R. M. (2020). AI4SAFE-IoT: An AI-powered secure architecture for edge layer of Internet of Things. Neural Computing and Applications, 32(20), 16119-16133.

[8] Hussain, A. (2024). AI-Powered Solutions for Cloud Security: Ensuring HIPAA and SOX Compliance Through Secure Data Pipelines and Robust Network Protection.

[9] John, B. (2025). A Comprehensive Study on Security Challenges and Solutions in AI-Driven Cloud Platforms.

[10] Joseph, A. (2024). AI-Driven Cloud Security: Proactive Defense Against Evolving Cyber Threats. International Journal of Computer and Information Engineering, 18(5), 261-265.

[11] Juttukonda, S. (2024). AI-Driven Innovations, Cloud Architectures, and Data Security: A Comprehensive Review of Emerging Technologies Across Domains.

[12] Kanth, T. C. (2024). AI-Powered Threat Intelligence for Proactive Security Monitoring in Cloud Infrastructures.

[13] Khan, M. M. (2024). Developing AI-Powered Intrusion Detection System for Cloud Infrastructure. Journal of Artificial Intelligence, Machine Learning and Data Science, 2(1), 1074-1080.

[14] Laura, M., & James, A. (2019). Cloud Security Mastery: Integrating Firewalls and AI-Powered Defenses for Enterprise Protection. International Journal of Trend in Scientific Research and Development, 3(3), 2000-2007.

[15] Mallikarjunaradhya, V., Pothukuchi, A. S., & Kota, L. V. (2023). An Overview of the Strategic Advantages of AI-Powered Threat Intelligence in the Cloud. Journal of Science & Technology, 4(4), 1-12.

[16] Mazhar, N., Noman, M., & Tahir, F. (2020). A Survey of Cloud Security Architectures: From Traditional to AI-Driven Solutions. International Journal of Digital Innovation, 1(1).

[17] Naveen, K. K., Priya, V., Sunkad, R. G., & Pradeep, N. (2024). An Overview of Cloud Computing for Data-Driven Intelligent Systems with AI Services. Data-Driven Systems and Intelligent Applications, 72-118.

[18] Nayak, A., Patnaik, A., Satpathy, I., & Patnaik, B. C. M. (2024). Data Storage and Transmission Security in the Cloud: The Artificial Intelligence (AI) Edge.

[19] Paul, F. (2023). AI-Powered Threat Detection in Hybrid and Multi-Cloud Environments: Overcoming Security Challenges.

[20] Paul, F. (2023). The Future of Cloud Security: AI-Powered Predictive Analytics for Proactive Threat Management.

[21] Rajesh, S. C., & Borada, D. AI-Powered Solutions for Proactive Monitoring and Alerting in Cloud-Based Architectures.

[22] Reddy, A. R. P. (2022). The Future of Cloud Security: AI-Powered Threat Intelligence and Response. International Neurourology Journal, 26(4), 45-52.

[23] Rehan, H. (2023). AI-Powered Genomic Analysis in the Cloud: Enhancing Precision Medicine and Ensuring Data Security in Biomedical Research. Journal of Deep Learning in Genomic Data Analysis, 3(1), 37-71.

[24] Rehan, H. (2024). Revolutionizing America's Cloud Computing: The Pivotal Role of AI in Driving Innovation and Security. Journal of Artificial Intelligence General Science (JAIGS), 2(1), 239-240.

[25] Sahid, F., & Hussain, K. (2018). AI-Powered DevOps and DataOps: Shaping the Future of Enterprise Architecture in the Cloud Era.

[26] Segar, M., & Zolkipli, M. F. (2024). A Study On AI-Driven Solutions for Cloud Security Platform. INTI Journal, 2024.

[27] Vadlamani, S., Kankanampati, P. K., Agarwal, R., Jain, S., & Jain, A. (2024). Integrating Cloud-Based Data Architectures for Scalable Enterprise Solutions. International Journal of Electrical and Electronics Engineering, 13(1), 21-48.

[28] Venkatesan, K. Enhancing Cybersecurity for National Infrastructure Through AI-Powered Cloud Monitoring Systems.

[29] Wang, J. (2023). AI/ML-Powered Cybersecurity and Cloud Computing Strategies for Optimized Business Intelligence in ERP Cloud.

[30] Wu, Y. (2020). Cloud-Edge Orchestration for the Internet of Things: Architecture and AI-Powered Data Processing. IEEE Internet of Things Journal, 8(16), 12792-12805.

**Declaration of interests**

☐ The authors declare that they have no known competing financial interests or personal relationships that could have appeared to influence the work reported in this paper.

☐The authors declare the following financial interests/personal relationships which may be considered as potential competing interests:

|  |
|---|
|  |

☐ XX is an editorial board member/associate editor/editor-in-chief for Cyber Security and Applications and was not involved in the editorial review or the decision to publish this article. All authors declare that there are no competing interests.